\documentclass[onecolumn,preprint,pre,12pt]{revtex4}
\usepackage{bm,epsf}

\begin{document}

\title{Inconsistency of a dissipative contribution to the mass flux in hydrodynamics}
\author{Hans Christian \"Ottinger}
\affiliation{Department of Materials, ETH Z\"urich, HCI H 543, CH-8093 Z\"urich,
Switzerland}
\author{Henning Struchtrup}
\affiliation{Department of Mechanical Engineering, University of Victoria, PO Box 3055
STN CSC, Victoria BC V8W 3P6, Canada}
\author{Mario Liu}
\affiliation{Theoretische Physik, Universit\"at T\"ubingen, D-72076 T\"ubingen, Germany}
\date{\today }

\begin{abstract}
The possibility of dissipative contributions to the mass flux is considered in detail. A general, thermodynamically consistent framework is developed to obtain such terms, the compatibility of which with general principles is then checked---including Galilean invariance, the possibility of steady rigid rotation and uniform center-of-mass motion, the existence of a locally conserved angular momentum, and material objectivity. All previously discussed scenarios of dissipative mass fluxes are found to be ruled out by some combinations of these principles, but not a new one that includes a smoothed velocity field $\bar{v}$. However, this field $\bar{v}$ is nonlocal and leads to unacceptable consequences in specific situations. Hence we can state with confidence that a dissipative contribution to the mass flux is not possible.
\end{abstract}

\maketitle

\section{The problem and its history}

The equality of mass flux and momentum density is widely accepted as a
statement of plausibility in hydrodynamics, supported by countless
experiments. However, in a classical paper, Dzyaloshinskii and Volovick \cite%
{DzyalVolov80} proposed the inclusion of a dissipative mass flux into the
hydrodynamic equations. Starting from a modified kinetic theory,
Klimontovich \cite{Klimontovich92} arrived at the same suggestion. More
recently, a dissipative contribution to the mass flux was re-introduced and
forcefully promoted by Brenner \cite{Brenner05b,Brenner06}, whose work
stimulated significant interest and controversy in the physics and fluid
dynamics communities.

In the early work of Brenner \cite{Brenner05b}, the mass flux and the
momentum density are taken to be equal, and only the volume velocity, used
in Newton's expression for the stress tensor and in no-slip boundary
conditions, differs by a dissipative term from the mass velocity (see
also Brenner's most recent work \cite{Brenner09}). Brenner's
idea has been expanded in \cite{ChakrabDurst07}, where not only Newton's
expression for the stress tensor but also Fourier's law for the heat flux
has been modified. These methods yield some elements, e.g. thermal stresses,
that are known in kinetic theory of gases as rarefaction effects which only
play a role beyond the hydrodynamic regime, that is, at finite Knudsen numbers.

Indeed, extensions of hydrodynamics to the finite Knudsen numbers are widely
discussed within kinetic theory (e.g., Burnett equations, R13 equations,
extended thermodynamics \cite{Struchtrup}); none of these includes
dissipative contributions to mass flux, and all reduce to classical
hydrodynamics in the limit of small Knudsen numbers. The modifications to
hydrodynamics in \cite{Brenner05b,ChakrabDurst07} are not able to produce
the whole wealth of rarefaction effects known in kinetic theory. Moreover,\
to our knowledge, they were never tested successfully against solutions of
the Boltzmann equation, or experiments.

Greenshields and Reese \cite{GreenshieldsReese07,GreenshieldsReese07ip} used
Brenner's equations of both scenarios \cite{Brenner05b,Brenner06} to
describe experimental findings for the structure of shock waves. In their
work, the modified hydrodynamic equations are actually considered as the
simplest way of introducing finite Knudsen number effects \cite{Struchtrup}.
However, the equations should be subjected to a much wider array of test
problems at small and finite Knudsen numbers, before one can claim that they
give meaningful results. Because we are interested in hydrodynamic theories
introducing a difference between mass flux and momentum density, however, we
do not discuss the ideas of \cite{Brenner05b,ChakrabDurst07} any further.

In an unpublished precursor version of \cite{Brenner05b}, dated October 30,
2002, Brenner had actually proposed a difference between mass flux and
momentum density, which is the topic of the present work. He later returned
to this original idea in \cite{Brenner06}, with a specific recommendation
for the cross coupling of density and temperature effects based on his
``incompressibility hypothesis.''

Brenner's work motivated thorough investigations on the thermodynamic
admissibility of a dissipative contribution to the mass flux \cite%
{hcobet,hco166,hco164,hco170,hco175}, which demonstrated that the idea fits
naturally into the GENERIC framework (``general equation for the
nonequilibrium reversible-irreversible coupling'') and into standard linear
irreversible thermodynamics. However, these investigations focused entirely
on nonequilibrium thermodynamics and neglected other, equally important
considerations, such as the local conservation of angular momentum. Such
additional criteria were considered earlier in \cite{KostadtLiu98} and
corroborated the original (not rigorously justified) statement of Landau and
Lifshitz that mass flux and momentum density must be equal (see footnote at
the end of Sec.~49 of \cite{LandauLifshitz6}). A concise summary of the
current state of the discussion can be found in the comment \cite{Liu08} to
the letter \cite{hco175}, and in the reply \cite{hco181} to that comment.

To obtain a more complete picture, we here first develop a general
thermodynamically consistent scenario of hydrodynamics with dissipative mass
flux and then apply the following additional criteria: Galilean invariance,
possibility of steady rigid fluid rotation, existence of a locally conserved
angular momentum, and consistency with uniform center-of-mass motion. To
keep the present analysis as simple as possible, we restrict ourselves to
single component fluids, we assume the absence of external forces, and we
use a vanishing bulk viscosity. All these limitations can be removed without
changing the arguments or conclusions in any essential way.

In the following we shall show that a consistent system of hydrodynamic
equations with a difference between mass flux and momentum density is only
possible if an additional velocity, denoted as $\bar{\bm{v}}$, is
introduced. While this velocity first arises as a modeling possibility--a
free parameter--within the GENERIC framework, it turns out that the criteria
listed in the previous paragraph can only be fulfilled for non-vanishing and
homogeneous $\bar{\bm{v}}$. However, we shall argue that a homogeneous
velocity with these properties is non-physical. We conclude that dissipative
contributions to mass flux are forbidden by the intricate interplay of the
various physical requirements. Previous discussions did not consider the
full breadth of physical criteria, and thus could not completely exclude
dissipative mass flow.

\section{Development of modified hydrodynamic equations}

To provide a reliable starting point for our discussion, we first introduce
a generalization of a previously developed thermodynamically consistent
scenario for hydrodynamic equations with a dissipative contribution to the
mass flux within the GENERIC framework of nonequilibrium thermodynamics.
Because we are concerned with the mass flux and the momentum density, the
key equations are the continuity equation for the mass density $\rho $ and
the equation of motion for the momentum density $\bm{M}$. To elucidate
thermodynamic aspects and to identify dissipative fluxes unambiguously, we
add the balance equation for the entropy density $s$ to obtain a complete
set of hydrodynamic equations, so that our final set of equations is for the
variables $x^{\prime }=(\rho ,\bm{M},s)$. For the formulation of the
friction matrix, however, it is more convenient to work with the variables $%
x=(\rho ,\bm{M},\epsilon )$, where $\epsilon $ is the internal energy
density.

For the formulation of classical hydrodynamics within the GENERIC framework
of nonequilibrium thermodynamics, we refer the reader to Section~2.2 of \cite%
{hcobet}. The generators energy and entropy are given there, together with
the Poisson and friction matrices turning the gradients of energy and
entropy into the reversible and irreversible contributions to time
evolution, respectively. We here restrict ourselves to the formulation of a
dissipative contribution to the mass flux, which is achieved through an
additional contribution to the friction matrix. Compared to the diffusive
contribution $M^{\mathrm{diff}}$ to the friction matrix given in Eq.~(2.77)
of \cite{hcobet}, we here consider a more general choice of a factorized
form of rank unity,
\begin{equation}
M^{\mathrm{diff}}=C\cdot D^{\prime }C^{T},  \label{Malternative}
\end{equation}%
where $D^{\prime }\geq 0$ is a transport coefficient associated with
diffusion, and the column vector $C$ and its transpose will be discussed below.

The dissipative contribution to the transport equations is given by (see %
\cite{hcobet}, Eqs. (1.1) and (2.54))%
\[
M^{\mathrm{diff}}\cdot \frac{\delta S}{\delta x}=M^{\mathrm{diff}}\cdot
\left(
\begin{array}{c}
-\frac{\mu }{T} \\
0 \\
\frac{1}{T}%
\end{array}%
\right)
\]%
where $S$ denotes the entropy of the system, and $\mu $ is the chemical
potential; $\frac{\delta }{\delta x}$ denotes a functional derivative. The
friction matrix $M^{\mathrm{diff}}$ has to fulfill the requirement of
degeneracy of energy $E$ (see \cite{hcobet}, Eqs. (1.5) and (2.49))
\begin{equation}
M^{\mathrm{diff}}\cdot \frac{\delta E}{\delta x}=M^{\mathrm{diff}}\cdot
\left(
\begin{array}{c}
-\frac{1}{2}\bm{v}^{2} \\
\bm{v} \\
1%
\end{array}%
\right) =0\;,  \label{Edegexpl}
\end{equation}%
which requires $C^{T}\cdot \frac{\delta E}{\delta x}=0$.

The elements of $C$ are a vector, a tensor, and a vector. To construct $C$
we assume that its elements must be tensorial combinations of $\left\{ \frac{%
\partial }{\partial \bm{r}},\bm{v},\bar{\bm{v}}\right\} $, where $\bm{v}$
denotes the velocity associated with momentum, and $\bar{\bm{v}}$ denotes
another velocity, which is introduced to account for possible differences
between mass velocity and momentum velocity, as stated in the introduction.

The elements of the column vector $C$ are subject to the following restrictions:\

(a) The dissipative contribution to the transport equations, $M^{\mathrm{diff%
}}\cdot \frac{\delta S}{\delta x}$, contains only second order space
derivatives, so that the additional dissipative contributions have the same
structure as the classical dissipative contributions from the laws of
Navier-Stokes and Fourier; thus, $C$ must be of first order in $\frac{%
\partial }{\partial \bm{r}}$.

(b) The contributions to mass and momentum balance must be of divergence
form, so that these have the proper form of conservation laws; thus, the
gradient must be on the left in the first and second term.

(c) $C$ is at most quadratic in the velocities. Due to the degeneracy requirement (\ref{Edegexpl}) this implies that the first two elements of $C$ must at most be linear in the velocities.

From the above follows immediately, that $C^{\left( 1\right) }=\frac{%
\partial }{\partial \bm{r}}$, which implies $C^{T\left( 1\right) }=-\frac{%
\partial }{\partial \bm{r}}$.

At first glance, the second element could be of the general form $C^{\left(
2\right) }=(\frac{\partial }{\partial \bm{r}}a_{1}\bar{\bm{v}})^{T}+(\frac{%
\partial }{\partial \bm{r}}a_{2}\bm{v})^{T}$. We anticipate the requirements
of Galilean invariance. Galilean invariance of the momentum balance is only
fulfilled for $a_{1}+a_{2}=1$, and since the velocity $\bar{\bm{v}}$ is
undefined yet, we can redefine as$\ a_{1}\bar{\bm{v}}+\left( 1-a_{1}\right) %
\bm{v}\rightarrow \bar{\bm{v}}$. Thus, we have $C^{\left( 2\right) }=(\frac{%
\partial }{\partial \bm{r}}\bar{\bm{v}})^{T}$ and $C^{T\left( 2\right) }=-(%
\bar{\bm{v}}\frac{\partial }{\partial \bm{r}})^{T}$.

The third component follows from the degeneration of energy as $C^{T\left(
3\right) }=\frac{\partial }{\partial \bm{r}}\frac{1}{2}\bm{v}^{2}-\bar{\bm{v}%
}\cdot (\frac{\partial }{\partial \bm{r}}\bm{v})^{T}+\hat{\alpha}\frac{%
\partial }{\partial \bm{r}}$. Here, $\hat{\alpha}$ is a scalar quantity
which possibly can depend on the velocities. Galilean invariance of the
dissipative mass flux requires that $\left( \hat{\alpha}-\frac{1}{2}\bar{v}%
^{2}\right) $ is Galilei invariant, so that we must have
\[
\hat{\alpha}-\frac{1}{2}\bar{v}^{2}=\bar{\alpha}+\beta \left( v-\bar{v}%
\right) ^{2}=\alpha
\]%
where $\bar{\alpha}$, $\beta $ are velocity independent scalars, and $\alpha
$ is a velocity dependent Galilei invariant scalar.

There actually is the possibility to add a further term of the form $%
(\partial /\partial \bm{r})(\hat{\bm{v}}-\bm{v})$ to $C^{\left( 2\right) }$,
without transposition, and a compensating term $(\hat{\bm{v}}-\bm{v})\,%
\mathrm{div}\bm{v}$ to $C^{\left( 3\right) }$. This ansatz involves a
further velocity field $\hat{\bm{v}}$. The resulting additional pressure
tensor contribution turns out to be isotropic and proportional to $\mathrm{%
div}\bm{v}$. In the following, we do not further consider such a
bulk-viscosity type effect by choosing $\hat{\bm{v}}=\bm{v}$.

Summarizing, we have the following%
\begin{equation}
C=\left(
\begin{array}{c}
\frac{\partial }{\partial \bm{r}} \\
(\frac{\partial }{\partial \bm{r}}\bar{\bm{v}})^{T} \\
\frac{1}{2}\bm{v}^{2}\frac{\partial }{\partial \bm{r}}-(\bm{v}\frac{\partial
}{\partial \bm{r}})^{T}\cdot \bar{\bm{v}}+\frac{\partial }{\partial \bm{r}}(%
\frac{1}{2}\bar{\bm{v}}^{2}+\alpha )%
\end{array}%
\right) ,  \label{Malternative1}
\end{equation}%
and
\begin{equation}
C^{T}=-\left(
\begin{array}{ccc}
\frac{\partial }{\partial \bm{r}} & \qquad (\bar{\bm{v}}\frac{\partial }{%
\partial \bm{r}})^{T}\qquad & \frac{\partial }{\partial \bm{r}}\frac{1}{2}%
\bm{v}^{2}-\bar{\bm{v}}\cdot (\frac{\partial }{\partial \bm{r}}\bm{v})^{T}+(%
\frac{1}{2}\bar{\bm{v}}^{2}+\alpha )\frac{\partial }{\partial \bm{r}}%
\end{array}%
\right) .  \label{Malternative2}
\end{equation}%
In these equations, the velocity field $\bm{v}=\bm{M}/\rho $ occurs as the
functional derivative of the energy with respect to the momentum density
field, whereas $\bar{\bm{v}}$ and $\alpha $ appear as free parameters in the
most general form of the friction matrix and need to be interpreted
according to their respective role in the hydrodynamic equations. For $\bar{%
\bm{v}}=\bm{v}$, one recovers the previously suggested scenario formulated
in \cite{hcobet}. The parameter $\alpha$ implies a
cross effect describing the influence of a temperature gradient on the
diffusive mass flux. We refer to $\bar{\bm{v}}$ as a smoothed velocity
field. As will be seen in the course of our arguments, non-vanishing $\bar{%
\bm{v}}$\ is necessary to obtain all required properties of the modified
hydrodynamic equations. The question whether such a velocity exists will be
discussed in our conclusions (Section VIII) based on the various
restrictions that will become apparent in the course of the paper.

The generalized contribution $M^{\mathrm{diff}}$ defined in Eqs.~(\ref%
{Malternative})--(\ref{Malternative2}) has all the properties required by
the thermodynamic framework: by construction $M^{\mathrm{diff}}$ is
symmetric, positive-semidefinite, and fulfills the degeneracy of energy. The
rank of $M^{\mathrm{diff}}$ is chosen to be unity because $M^{\mathrm{diff}}$
represents a single additional dissipative process.

For the generalized friction matrix with the contribution (\ref{Malternative}%
) due to diffusion, after transforming from the variables $x$ to $x^{\prime
} $, we obtain a hydrodynamic scenario expressed by the following equations:
\begin{equation}
\frac{\partial \rho }{\partial t}=-\frac{\partial }{\partial \bm{r}}\cdot (%
\bm{M}-\bm{j}\rho ),  \label{rhoconteqmod}
\end{equation}%
\begin{equation}
\frac{\partial \bm{M}}{\partial t}=-\frac{\partial }{\partial \bm{r}}\cdot
\left( \frac{1}{\rho }\bm{M}\bm{M}-\bm{j}\rho \bar{\bm{v}}+p\bm{1}+\bm{\tau}%
\right) ,  \label{uNavStoeqmod}
\end{equation}%
and
\begin{equation}
\frac{\partial s}{\partial t}=-\frac{\partial }{\partial \bm{r}}\cdot \left(
\frac{1}{\rho }\bm{M}s+\bm{j}^{s}\right) +\frac{\rho ^{2}\bm{j}^{2}}{%
D^{\prime }}+\frac{{\bm{j}^{\mathrm{q}}}^{2}}{\lambda ^{\mathrm{q}}T^{2}}+%
\frac{\bm{\tau}:\bm{\tau}}{2\eta T}.  \label{entropevoleqmod}
\end{equation}%
For completeness, we also give the energy balance, which can be obtained
from combining mass and entropy balance via the Gibbs equation,
\begin{equation}
\frac{\partial u}{\partial t}=-\frac{\partial }{\partial \bm{r}}\cdot \left(
\frac{\mathbf{M}}{\rho }u+\bm{j}^{\mathrm{e}}\right) +\left( -p\bm{1}+%
\bm{\tau}\right) :\frac{\bm{\tau}}{2\eta }+\rho \left( \bar{\bm{v}}-\bm{v}%
\right) \cdot \frac{\partial \bm{v}}{\partial \bm{r}}\cdot \bm{j}\;.
\label{energy balance}
\end{equation}

On the right-hand sides of these equations, there occur the local
equilibrium pressure $p=p(\rho ,s)$, as well as the dissipative fluxes of
mass ($\rho \bm{j}$), momentum ($\bm{\tau}$), heat ($\bm{j}^{\mathrm{q}}$),
entropy ($\bm{j}^{s}$), and energy ($\bm{j}^{\mathrm{e}}$).

These fluxes are given in terms of transport coefficients and gradients of
intensive quantities:
\begin{equation}
\bm{j}^{\mathrm{q}}=\lambda ^{\mathrm{q}}T^{2}\frac{\partial }{\partial %
\bm{r}}\frac{1}{T}=-\lambda ^{\mathrm{q}}\frac{\partial T}{\partial \bm{r}}
\label{lambdaFour}
\end{equation}%
is the thermal contribution to non-convective energy transport, in terms of
the thermal-conductivity parameter $\lambda ^{\mathrm{q}}$ and the gradient
of the absolute temperature $T$. The complete non-convective energy flux is
given by
\[
\bm{j}^{\mathrm{e}}=\bm{j}^{\mathrm{q}}-\rho \bm{j}\left( \alpha +\frac{1}{2}%
\left( v-\bar{v}\right) ^{2}\right) \;;
\]%
clearly, $\alpha $ describes a cross effect. Furthermore,
\begin{equation}
\bm{\tau}=-\eta \left[ \frac{\partial }{\partial \bm{r}}\bm{v}+\left( \frac{%
\partial }{\partial \bm{r}}\bm{v}\right) ^{T}-\frac{2}{3}\frac{\partial }{%
\partial \bm{r}}\cdot \bm{v}\bm{1}\right]  \label{tauNewt}
\end{equation}%
is Newton's expression for the stress tensor in terms of the viscosity $\eta
$ and the gradient of the velocity (as mentioned before, it is convenient to
assume that the bulk viscosity vanishes), and
\begin{equation}
\rho \bm{j}= D^{\prime } \left[ \frac{\partial }{\partial \bm{r}}%
\frac{\tilde{\mu}}{T}-\alpha \frac{\partial }{\partial \bm{r}}\frac{1}{T}-%
\frac{1}{T}(\bm{v}-\bar{\bm{v}})\cdot \left( \frac{\partial }{\partial \bm{r}%
}\bar{\bm{v}}\right) ^{T}\right]  \label{massflux}
\end{equation}%
is the dissipative mass flux in terms of the coefficient $D^{\prime }$,
which is closely related to the diffusion coefficient, the cross coefficient
$\alpha $, and the gradient of a velocity-modified version of the chemical
potential per unit mass $\mu $ in addition to the gradient of temperature
\begin{equation}
\tilde{\mu}=\mu -\frac{1}{2}(\bm{v}-\bar{\bm{v}})^{2}\;;  \label{mumod}
\end{equation}%
the total mass flux is given by $(\bm{M}-\bm{j}\rho )$.

Finally,
\begin{equation}
\bm{j}^{s}=\frac{1}{T}\bm{j}^{\mathrm{q}}+\frac{\tilde{\mu}-\alpha}{T%
}\rho\bm{j}=\frac{1}{T}\bm{j}^{\mathrm{e}}+\frac{\mu }{T} \rho\bm{j}
\label{entropflux}
\end{equation}%
is the total entropy flux; the second equation shows the expected form of
the entropy flux for systems with energy and diffusion fluxes.

There occurs a natural quadratic entropy production term associated with the
dissipative mass flux $\rho\bm{j}$ (that is, with diffusion), so that the
entropy production remains positive semidefinite. Equation (\ref%
{entropevoleqmod}) suggests that the added diffusion effect is on the same
footing as the dissipative effects in conventional hydrodynamics.

The natural occurrence of a velocity-dependent chemical potential has
previously been found in \cite{KostadtLiu98} (however, without $\bar{\bm{v}}$%
). Such a modification of the chemical potential suggests that a possible
diffusion process is more subtle than viscous stresses or thermal
conductivity. Note that the generalizations occur only through second-order
terms in the deviations from equilibrium. In linearized hydrodynamics, $\bar{%
\bm{v}}$ does not play any role.

From a conceptual perspective, it is important to notice that we need to
introduce two different velocities $\bm{v}$ and $\bar{\bm{v}}$ before we can
formulate a dissipative contribution to the mass flux. Therefore, the
discrepancy between mass flux and momentum density is not the primary reason
for the occurrence of two different velocities; rather two different
velocities are needed to introduce diffusion into hydrodynamics. Unlike
momentum velocity $\bm{v}=\frac{\bm{M}}{\rho }$\ and mass velocity $\frac{%
\bm{M}}{\rho }-\bm{j}$, the additional velocity $\bar{\bm{v}}$\ does not
have an obvious intuitive interpretation; as the equations show, non-zero $\bar{%
\bm{v}}$\ would manifest itself through its contribution to diffusion.

\section{Galilean invariance}

Galilean invariance was already used in the construction of the friction
matrix $M^{\mathrm{diff}}$, although we did not present explicit arguments
to shorten the presentation.

If we introduce $\bm{v}_{0}$ as an arbitrary constant velocity shift, and
make the replacements
\begin{equation}
\bm{M}\rightarrow \bm{M}+\rho \bm{v}_{0},\qquad \bar{\bm{v}}\rightarrow \bar{%
\bm{v}}+\bm{v}_{0},\qquad \frac{\partial }{\partial t}\rightarrow \frac{%
\partial }{\partial t}-\bm{v}_{0}\cdot \frac{\partial }{\partial \bm{r}},
\label{GIrep}
\end{equation}%
in the hydrodynamic scenario given by the mass balance (\ref{rhoconteqmod}),
the momentum balance (\ref{uNavStoeqmod}), and the entropy balance (\ref%
{entropevoleqmod}), they keep their original form. For this to be true it is
essential that $\mu $ and $\tilde{\mu}$ and hence also $\bm{j}$ and $\bm{j}%
^{s}$ are invariant under Galilean transformations, and also the
cross coefficient $\alpha $ needs to be invariant. As indicated in the
previous section, the requirement of invariance already was considered in
the construction of the friction matrix. We note that invariance of the
transport equations does not require invariance of the friction matrix.

While Galilean invariance concerns the transformation between inertial
systems, one can also consider the question of transformations into
non-inertial frames, which in classical hydrodynamics leads to the occurrence
of inertial forces \cite{IngoMuller85}. In the present context of modified
hydrodynamics with diffusive mass flow, there will be inertial terms in the
mass flux as well, with subtle consequences for material objectivity \cite%
{IngoMuller85}. Our line of arguments does not require discussion of
non-inertial frames, and will lead to the conclusion that diffusive
contributions to mass flow are not possible; hence we refrain from a more
detailed discussion of these issues.

\section{Integrability}

Subtracting $\left( -\frac{\partial \bm{j}\rho }{\partial t}\right) $ from
both sides of the momentum balance (\ref{uNavStoeqmod}) gives the transport
equation for the total mass flux $\left( \bm{M}-\bm{j}\rho \right) $. This
equation is a conservation law if a suitable function $J$ exits as a
potential for $\bm{j}\rho$, so that
\begin{equation}
\bm{j}\rho =\frac{\partial J}{\partial \bm{r}}\;.  \label{integrability}
\end{equation}%
In general, $J$ could be a tensor, but in the following we consider only
scalar functions. Note that the existence of a potential $J$ is sufficient
for local conservation of the mass flux, but not necessary, that is, the mass flux
could be conserved under weaker conditions. For the further arguments, however,
a detailed discussion of the condition (\ref{integrability}) turns out to be useful.

For constant (i.e., homogeneous) parameters $D^{\prime }$, $\alpha $, and $\bar{\bm{v}}$ in Eqs.~(\ref{massflux}), the dissipative contribution to the mass flux can trivially be written in the form (\ref{integrability}). However, such a representation holds under much weaker conditions on $D^{\prime }$, $\alpha $, and $\bar{\bm{v}}$, and that is our motivation for proposing it. In the following, we refer to Eq.~(\ref{integrability}) as the \emph{integrability condition}.

A local conservation law for the mass flux is, in fact, expected from
general considerations of statistical mechanics \cite{EvansMorriss}, and it
probably is the reason why one usually concludes without hesitation that the
mass flux must coincide with the well-known conserved momentum density.
Below it will be shown that a second conservation law originates in the
condition (\ref{integrability}) which, in turn, is needed to obtain a
uniform center-of-mass motion.

We consider the most general form for $J$\ as a function of the hydrodynamic
variables up to second order terms in velocity, which reads\textbf{\ }
\begin{equation}
J=J_{1}(\mu ,T)-\frac{1}{2}J_{2}(\mu ,T)(\bm{v}-\bm{c})^{2},  \label{Jansatz}
\end{equation}%
Galilean invariance of $J$\ requires the dependence on a velocity
difference, hence the appearance of the additional velocity $\bm{c}$, which
will be determined by comparing Eqs.~(\ref{massflux}) and (\ref%
{integrability}) for the ansatz (\ref{Jansatz}). We find right away that%
\textbf{\ }%
\begin{equation}
\frac{\partial \bm{c}}{\partial \bm{r}}=0\;,  \label{cconst}
\end{equation}%
that is the velocity $\bm{c}$\ must be homogeneous (constant in space).

Next, we identify the diffusion coefficient as
\begin{equation}
\frac{D^{\prime }}{T}=\frac{\partial J_{1}}{\partial \mu }-\frac{1}{2}\frac{%
\partial J_{2}}{\partial \mu }(\bm{v}-\bm{c})^{2}\;.  \label{J1choice}
\end{equation}%
Thus, an arbitrary functional form of the diffusion coefficient can be
reproduced by choosing a suitable $\mu $-dependence of $J_{1}$ and $J_{2}$.
In fact, to restrict the constitutive law (\ref{massflux}) to second order
in velocity, one will choose $\frac{\partial J_{2}}{\partial \mu }=0$, so
that $J_{2}=J_{2}\left( T\right) $.

By comparing the remaining terms, we further obtain
\begin{equation}
\bar{\bm{v}}=\bm{v}+\frac{TJ_{2}(T)}{D^{\prime }}(\bm{c}-\bm{v}),
\label{vbarexp}
\end{equation}%
and
\begin{equation}
\alpha =\mu +\frac{T^{2}}{D^{\prime }}\,\frac{\partial J_{1}}{\partial T}-%
\frac{1}{2}\left( 1-D^{\prime }\frac{\partial }{\partial T}\frac{1}{J_{2}}%
\right) (\bm{v}-\bar{\bm{v}})^{2}.  \label{alexp}
\end{equation}

We consider special choices of $J_{2}$, and their influence on the
constitutive equation (\ref{vbarexp}) for $\bar{\bm{v}}$. Note that the
previously suggested choice $J_{2}=0$ \cite{hcobet}\ gives $\bar{\bm{v}}=%
\bm{v}$\ by Eq.~(\ref{vbarexp}) in agreement with \cite{hcobet}. If $%
D^{\prime }$ depends only on temperature, the natural choice $%
J_{2}=D^{\prime }\left( T\right) /T$ leads to the simple expression $\bar{%
\bm{v}}=\bm{c}$; in this case, because of (\ref{cconst}), $\bar{\bm{v}}$
must be a constant. However, if $D^{\prime }$ depends also on $\mu $, the
reference velocity $\bar{\bm{v}}$ cannot just be a constant.

With the assumption that for an ideal gas the mass diffusion coefficient $%
D^{\prime }$ behaves similar to the self-diffusion coefficient, $D^{\prime
}\propto \sqrt{T}$ depends only on $T$, so that $J_{1}=\mu D^{\prime }/T$
and $J_{2}=D^{\prime }/T$ are the most natural choices. We hence find
\begin{equation}
\bar{\bm{v}}=\bm{c},\qquad \alpha =\frac{1}{2}\tilde{\mu}.
\label{idgasalpha}
\end{equation}%
The example of the ideal gas demonstrates that the integrability condition
can indeed be consistent with the generalized hydrodynamic scenario. The
expression (\ref{idgasalpha}) for $\alpha $\ differs from the values
considered in the pertinent literature \cite{Brenner06,hcobet}. While one
might argue that the above result for $\alpha $\ for an ideal gas casts the
idea of integrability into doubt, one should take into account that all
suggestions for non-zero values of $\alpha $ and $D^{\prime }$ (and thus
non-zero dissipative contribution to mass flux) are based solely on
plausibility arguments, but not on derivation from first principles nor on
experimental evidence. Note also that, under the assumption of integrability,
the above discussion suggests that $%
\bar{\bm{v}}$\ is homogeneous (at least for the ideal gas)!

\section{Angular momentum}

In any hydrodynamic theory, one should be able to construct a locally
conserved angular momentum resulting from the existence of rotational
symmetry \cite{Goldstein}. The conservation or non-conservation of angular
momentum is related to the structure of the momentum balance equation. When
the latter is taken into the cross product with $\bm{r}$, we find a balance
for the local angular momentum density $\bm{l}=\bm{r}\times \bm{M}$, which
we write in index notation as
\begin{equation}
\frac{\partial l_{i}}{\partial t}=-\frac{\partial }{\partial r_{l}}\left[
\frac{1}{\rho }M_{l}l_{i}+\epsilon _{ijk}r_{j}\left( -j_{l}\rho \bar{v}%
_{k}+p\delta _{kl}+\tau _{kl}\right) \right] -\epsilon _{ijk}\rho j_{j}\bar{v%
}_{k}.
\end{equation}%
One finds $-\epsilon _{ijk}\rho j_{j}\bar{v}_{k}=(\bar{\bm{v}}\times \bm{j}%
\rho)_i $ as a local source of angular momentum. This contribution stems form
the non-symmetric contribution $-\bm{j}\rho \bar{\bm{v}}$ to the momentum
flux tensor, which therefore potentially causes a problem. With the identity
\begin{equation}
\epsilon _{ijk}\rho j_{j}\bar{v}_{k}=\frac{\partial }{\partial r_{l}}\left[
\epsilon _{ljk}\rho j_{j}\bar{v}_{k}r_{i}\right] +r_{i}\bar{v}_{k}\left(
\epsilon _{kjl}\frac{\partial \rho j_{j}}{\partial r_{l}}\right) -r_{i}\rho
j_{j}\left( \epsilon _{jkl}\frac{\partial \bar{v}_{k}}{\partial r_{l}}%
\right) \;,  \label{lflux}
\end{equation}%
it becomes clear that we can obtain a locally conserved angular momentum
provided that the last two terms in (\ref{lflux}) vanish, that is if (i) the
integrability condition (\ref{integrability}) holds, and (ii) the modified
velocity field $\bar{\bm{v}}$ is curl-free.\textbf{\ }

Clearly, condition (ii) is fulfilled if we take $\bar{\bm{v}}$ as
independent of position. For the previous choice, $\bar{\bm{v}}=\bm{v}$,
however, this condition will not be fulfilled, that is this choice would
lead to violation of conservation of angular momentum, and thus is not
physical!

Conservation of angular momentum together with the integrability requirement
of Section IV leads to a strong reduction of the possible constitutive
equations:\ Together with curl-free $\bar{\bm{v}}$\ the condition (\ref%
{vbarexp}) can only be fulfilled for arbitrary fields $T,\mu $, $\bm{v}$\ if
$D^{\prime }$\ is independent of $(\bm{v}-\bm{c})$. Because of (\ref%
{J1choice}), this implies that $J_{2}$\ must be independent of $\mu $, and
then vanishing curl of (\ref{vbarexp}) requires $D^{\prime }=TJ_{2}\left(
T\right) $, so that $D^{\prime }$\ must depend only on $T$. With this, (\ref%
{J1choice}) reduces to $D^{\prime }\left( T\right) =T\frac{\partial J_{1}}{%
\partial \mu }$ which implies $J_{1}\left( \mu ,T\right) =\frac{\mu }{T}%
D^{\prime }\left( T\right) $. Thus, integrability and conservation of
angular momentum together require $\bar{\bm{v}}=\bm{c}$, that is constant $%
\bar{\bm{v}}$. We found the same in Section IV for the ideal gas; the
present argument extends the findings to all\ materials as sufficient, but
not necessary conditions.

Can one construct a locally conserved angular momentum without assuming integrability? It is generally expected that a non-symmetric stress tensor is associated with some internal angular momentum. However, we did not succeed in constructing an internal angular momentum in such a way that the total angular momentum is conserved. A hydrodynamic theory without the possibility to construct a locally conserved angular momentum density field needs to be rejected. The only construction we could find is based on integrability and constant $\bar{\bm{v}}$ and hence rules out the previous formulations of modified hydrodynamics \cite{Brenner06,hcobet} with a dissipative contribution to the mass flux based on $\bar{\bm{v}}=\bm{v}$. Any alternative suggestion of a scenario with a dissipative mass flux must come with an explicit suggestion for the total angular momentum density.

\section{Steady rigid fluid rotation}

To obtain a further criterion for the admissibility of a dissipative
contribution to the mass flux, we look for a particular solution of the
different sets of hydrodynamic equations introduced in the preceding
sections. Namely, we expect that a fluid performing a rigid body rotation
should provide a solution of any valid set of hydrodynamic equations. The
existence of dissipation free rigid rotations is not a necessary condition.
If one believes that such solutions should exist, however, one finds additional
restrictions on modifications of transport processes.

In particular, for steady rigid fluid rotation, all dissipative flux
contributions must vanish. Moreover, the pressure must increase radially to
provide the required centripetal forces,
\begin{equation}
\frac{dp}{dr}=\rho \frac{\bm{v}^{2}}{r},  \label{pressurecond}
\end{equation}%
where $r$ is the distance from the axis of rotation.

For the further arguments, it is important that thermodynamic local
equilibrium states are fully characterized by two intensive state variables.
Because the pressure $p$ must have the spatial variation (\ref{pressurecond}%
), at most one intensive local-equilibrium variable can be constant
throughout the system.

We first consider the case where $\bar{\bm{v}}=\bm{v}$, as was suggested for
the original modifications to hydrodynamics \cite{Brenner06,hcobet}. For
this case, according to Eqs.~(\ref{lambdaFour}) and (\ref{massflux}),
vanishing dissipative fluxes $\bm{j}^{\mathrm{q}}=\bm{j}=\bm{0}$ would
require that the two intensive variables $T$ and $\mu $ be constant
throughout the system, which is at variance with the spatial variation of $p$
given in Eq.~(\ref{pressurecond}). This implies that, for the previously
proposed formulation, there can only be a single process, either heat
conduction (that is, standard hydrodynamics), or just the combined effect of
diffusion and heat conduction described by $\bm{j}$. In other words, two
dissipative processes driven by gradients of two intensive local-equilibrium
variables, as assumed in all previous papers \cite%
{Klimontovich92,Brenner06,hcobet, hco166,hco164,GreenshieldsReese07ip}, are
excluded by the assumption of a solution describing steady rigid fluid
rotation\textbf{.}

For a single combined Brenner-type process, one expects to find a nontrivial temperature profile for the rigidly rotating fluid. This is so, because the pressure profile is dictated by the centripetal forces, and the combination of pressure and temperature gradients that drives the mixed process must vanish in the dissipation-free state of rotation. Note, however, that the existence of such a solution for a rigidly rotating fluid does not guarantee the general conservation of angular momentum.

For our more general formulation in the center-of-mass system, for which we
assume $\bar{\bm{v}}=\bm{0}$, according to Eqs.~(\ref{lambdaFour}), (\ref%
{massflux}) and (\ref{mumod}), vanishing dissipative fluxes $\bm{j}^{\mathrm{%
q}}=\bm{j}=\bm{0}$ would require only the intensive variable $T$ to be
constant throughout the system, whereas constant $\tilde{\mu}$ for rigid
rotation translates into
\begin{equation}
\frac{d\mu }{dr}=\frac{d}{dr}\left( \frac{1}{2}\bm{v}^{2}\right) =\frac{%
\bm{v}^{2}}{r}.  \label{mucond}
\end{equation}%
According to the Gibbs-Duhem relation, that is, $dp=sdT+\rho d\mu $, Eq.~(%
\ref{mucond}) is perfectly consistent with the radial pressure
distribution (\ref{pressurecond}) at constant temperature.

Whereas inhomogeneous equilibria caused by external forces can be described
in the previously suggested formulation because the external forces appear
in the thermodynamic driving forces of the dissipative processes \cite%
{hco166}, the inhomogeneous equilibrium due to fluid rotation can only be
handled through the velocity-modified chemical potential of the generalized
approach.

Note, however, that the above argument requires $\bar{\bm{v}}=\bm{0}$ in the
center-of-mass system, this condition must be seen as a severe constriction
for $\bar{\bm{v}}$. Integrability has not been assumed in this section.

\section{Center-of-mass position}

In the absence of external forces, the center-of-mass motion of a system
must be uniform. This fundamental insight allows us to reconstruct the
center-of-mass position at time $t=0$, which is a constant of motion. More
generally, one actually expects
\begin{equation}
\bm{b}=\rho \bm{r}-\bm{M}t  \label{boosterdenKL}
\end{equation}%
to be the density of a locally conserved quantity. From the perspective of
special relativity, this locally conserved quantity $\bm{b}$ is the natural
partner of the angular momentum density $\bm{l}$, and $\bm{b}$ is known as the booster
density \cite{KostadtLiu98}. From the definition (\ref{boosterdenKL}) and
the balance equations (\ref{rhoconteqmod}), (\ref{uNavStoeqmod}), we obtain%
\begin{equation}
\frac{\partial \bm{b}}{\partial t}=-\frac{\partial }{\partial \bm{r}}\cdot %
\left[ \frac{1}{\rho }\bm{M}\bm{b}-\bm{j}\rho (\bm{r}-\bar{\bm{v}}t)-t(p%
\bm{1}+\bm{\tau})\right] -\bm{j}\rho .  \label{boostereq}
\end{equation}%
Under the integrability condition (\ref{integrability}), the uniform
center-of-mass motion is associated with a locally conserved quantity $\bm{b}
$. Integrability, locally conserved angular momentum, locally conserved mass
flux, and locally conserved booster are hence intimately related features.
The generalized setting introduced in this paper provides a formally
acceptable modification of standard Navier-Stokes-Fourier hydrodynamics,
provided that we assume integrability.

Like for the angular momentum density $\bm{l}$, we might consider the
possibility to define the booster density $\bm{b}$ in terms of the mass flux
instead of the momentum density. Such a definition would seem plausible
because the local effects on the rearrangement of the center-of-mass
position are given by the mass flux. It then becomes evident that a local
conservation law for the modified booster is directly equivalent to a local
conservation law for the mass flux, without using integrability as an
intermediate link. There hence exists a deeper reason for a separate
conservation law for the mass flux.

\section{Conclusions and open problems}

In spite of its thermodynamic admissibility, the most widely discussed
scenario for introducing a dissipative contribution to the mass flux in
hydrodynamics \cite{Klimontovich92,Brenner06,hcobet,
hco166,hco164,GreenshieldsReese07ip} must be rejected because it violates
the local conservation of angular momentum. Moreover, it does not allow for
steady rigid fluid rotation. Through the occurrence of an additional
curl-free velocity $\bar{\bm{v}}$, the balance equations (\ref{rhoconteqmod}%
)--(\ref{entropevoleqmod}) together with the constitutive equations (\ref%
{lambdaFour})--(\ref{massflux}) provide a new generalized set of
hydrodynamic equations with a dissipative contribution to the mass flux. For
a suitable choice of the smoothed velocity field $\bar{\bm{v}}$, and with
the assumption of an integrability condition, all the following criteria can
be satisfied:

\begin{itemize}
\item Galilean invariance,

\item possibility of steady rigid fluid rotation,

\item existence of a locally conserved angular momentum,

\item and consistency with uniform center-of-mass motion.
\end{itemize}

In addition to the momentum density, also the mass flux turns out to be a
locally conserved quantity. The deeper reason for this conservation of the
mass flux lies in the existence of a locally conserved booster density. In
the new generalized formulation, an integrability condition ensures the
existence of the proper conservation laws.

Although it satisfies all the fundamental admissibility criteria, the new
generalized formulation of hydrodynamics must be scrutinized in view of the
role of the additional velocity field $\bar{\bm{v}}$. Assuming integrability,
we found that $\bar{%
\bm{v}}$\ must be curl-free to satisfy conservation of angular momentum
(Section V), that it must vanish in the center of mass of a rotating system
to allow rigid fluid rotation (Section VI), and that it must be constant to
guarantee conservation of mass flux (Sections IV/VI).

From this list, one might be inclined to assume that $\bar{\bm{v}}$ is
independent of position,\ that is, homogeneous, but possibly time-dependent,
and that the center-of-mass velocity is the simplest interpretation of $\bar{%
\bm{v}}$. However, this leads immediately to the question of how to define
the system for the determination of the center of mass. While for certain
systems a particular choice of boundaries might be self-suggesting, for example the
walls of a vessel, a local continuum theory must be valid for any choice of
system (provided that the system is bigger than a meaningful ``point'' of
the continuum).

As we look for a local theory of hydrodynamics, any given system can be
divided into smaller subsystems in an arbitrary manner, and this division
should not alter the mathematical description of the physics within the
system, or the subsystems. Thus, if $\bar{\bm{v}}$ is a homogeneous velocity
in a system $A$, it must be the same velocity in any subsystem of $A$. Also,
if $A$\ is combined with another system $B$, the combined system $A+B$\ must
still have the same $\bar{\bm{v}}$, since else the local description in $A$\
would be altered. As a logical consequence follows that $\bar{\bm{v}}$\
would have to be a universal constant. This is inconsistent with Galilean
invariance.

Only when the dissipative mass flux vanishes ($D^{\prime }=0$) do we have no
contradiction---but this is just the case of classical hydrodynamics. Thus,
we come to the conclusion that a dissipative mass flux necessarily leads to
contradiction to basic physical principles, and must be rejected. It follows
that classical hydrodynamics, where mass flux and momentum density agree, is
the only admissible system of transport equations in the hydrodynamic regime
(i.e., at small Knudsen numbers; for finite Knudsen numbers rarefaction
effects must be accounted for\ which leads to more complicated forms of the
transport equations which all reduce to classical hydrodynamics in the limit
of sufficiently small Knudsen numbers \cite{Struchtrup}).

A particular example arises by generalizing our example of steady rigid
fluid rotation. If a system consists of two independent, rigidly rotating
subsystems in relative motion, the two individual center-of-mass velocities
must be used to obtain a proper overall solution. The averaging is no longer
global but only over the subsystems. If, in general, the averaging required
to obtain the smoothed velocity field cannot be global, then the question
about the proper spatial averaging procedure arises. We are then faced with
mathematical and physical issues, and we here focus on the latter, in
particular: What is the relevant physical length scale for smoothing?

The fundamental problem associated with a dissipative contribution to the
mass flux is that it represents random fluctuations in the position of
non-inertial particles and hence destroys the uniform center-of-mass motion
implied by Newton's equations of motion. Only after some averaging procedure
one can expect a smooth center-of-mass motion. This averaging should at
least comprise several collisions, which define the intrinsic physical scale
of hydrodynamic equations. The collisions also define the characteristic
scales needed to reach local equilibrium. As the velocity $\bm{v}$ of
classical hydrodynamics should clearly be defined on the local-equilibrium
scale, there is no natural way of introducing a further smoothed velocity $%
\bar{\bm{v}}$ based on intrinsic length scales. Only if the hydrodynamic
velocity $\bm{v}$ is assumed to be well-defined on a more local scale \cite%
{AlderAlley84}, as implicitly done in fluctuating hydrodynamics (see Chapter
IX of Landau and Lifshitz \cite{LandauLifshitz9} or Section 2.2.4 of \cite%
{hcobet}) or in fluctuation renormalization (related to eliminating the
back-flow effects of hydrodynamics leading to long-time tails as described
in Chapter~16 of \cite{Reichl}), then there exists a natural smoothing
procedure. However, we then enter the domain of finite Knudsen number
effects and hence leave the scope of the purely hydrodynamic theory
discussed in this paper.

While a dissipative contribution to the mass flux seems to be irrelevant to
continuum hydrodynamics, a deeper understanding of the related phenomena may
still be useful to resolve some problems arising in the theory of
fluctuations \cite{hco175}. As a consequence, Einstein's theory of local
equilibrium fluctuations can only be valid on length scales large compared
to the mean free path, even for rarefied gases. More disturbingly, the use
of Onsager's regression hypothesis \cite{Reichl,OrtizSengers} to evaluate
the two-time correlations determining the dissipative properties of
nonequilibrium systems according to the Green-Kubo formula \cite%
{hcobet,Reichl,Grabert} needs to be questioned. However, it might also
happen that the problems related to the description of fluctuations in
hydrodynamics can only be resolved in a more complete consistent theory at
finite Knudsen numbers.

To summarize, we state that all known modifications of hydrodynamics that lead to a
dissipative contribution to mass flux violate physical principles and must
be discarded.

% \bibliography{hcopubs}

\expandafter\ifx\csname natexlab\endcsname\relax

\fi\expandafter\ifx\csname bibnamefont\endcsname\relax

\fi\expandafter\ifx\csname bibfnamefont\endcsname\relax

\fi\expandafter\ifx\csname citenamefont\endcsname\relax

\fi\expandafter\ifx\csname url\endcsname\relax

\fi\expandafter\ifx\csname urlprefix\endcsname\relax

\fi\providecommand{\bibinfo}[2]{#2} \providecommand{\eprint}[2][]{\url{#2}}

\end{document}